\documentclass[aip, graphicx]{revtex4-1}
\usepackage{graphicx,amsmath}
\usepackage{xcolor}
\usepackage{float}
\usepackage[english]{babel}

\draft % marks overfull lines with a black rule on the right

\begin{document}

%\preprint{}

\title{Contrasting ultrafast light-driven electron-hole interaction dynamics in monolayer MoS$_2$ and metallic NbSe$_2$} %Title of paper



\author{Aday C\'ardenas}
\affiliation{Instituto de Ciencia de Materiales de Madrid, CSIC, 28049  Madrid, Spain }%
\author{Rui E. F. Silva}
\affiliation{Instituto de Ciencia de Materiales de Madrid, CSIC, 28049  Madrid, Spain }%
\affiliation{Max Born Institute, Max-Born-Stra{\ss}e 2A, 12489, Berlin, Germany}%
\author{\'Alvaro Jim\'enez-Gal\'an}
\affiliation{Instituto de Ciencia de Materiales de Madrid, CSIC, 28049  Madrid, Spain }%
\affiliation{Max Born Institute, Max-Born-Stra{\ss}e 2A, 12489, Berlin, Germany}%
 \email{alvaro.jimenez@csic.es}


%\date{\today}

\begin{abstract}
We study strong-field driven ultrafast dynamics and high-harmonic generation (HHG) in monolayer 2H-NbSe$_2$ and compare them with those of monolayer 2H-MoS$_2$ by solving the multiband reduced-density-matrix equations including time-dependent electron-electron interaction effects within the time-dependent Hartree + screened exchange (TD-HSEX). In MoS$_2$, these interactions strongly enhance the harmonic yield and modify the harmonic phases and angular emission patterns, wheras in NbSe$_2$ the yield enhancement is weaker but clear phase and angular changes remain. We trace these differences to the distinct optical resonances and to the different bands involved in the emission in each material. Finally, we show that carrier injection into empty bands of NbSe$_2$ differs qualitatively from interband excitation in MoS$_2$, and is well captured at a qualitative level by a Keldysh tunneling rate with a time-dependent band separation, allowing to control the timing and the region of injection of carriers to empty bands of the metal with the field parameters. Our work provides a framework to interpret ultrafast electron-hole interaction effects in experimental high harmonic generation spectra across semiconducting and metallic systems.
\end{abstract}


\maketitle %\maketitle must follow title, authors, abstract and \pacs

\section{Introduction}
High-harmonic generation (HHG) has become a key time-resolved spectroscopic tool for atoms~\cite{LHuillierBalcou, MFerray_1988, LewensteinPRA24, KrauseSchaferKulander, KrauszIvanovRev}, molecules~\cite{smirnova_high_2009, silva_even_molecules_2016}, and solids~\cite{ghimire_high-harmonic_2019, VampaTutorial, heide_ultrafast_2024}. In the case of solids, many of their electronic properties are clearly imprinted on the harmonic spectrum, which has allowed for the reconstruction of the material band structure~\cite{PhysRevLett.115.193603}, probing of phase transitions~\cite{purschke2025giantenhancementattosecondtunnel, RuiHubbard}, observation of atomic-like orbital interference \cite{jimenez-galan_orbital_2023}, multi-band effects~\cite{Mette_Zuerch, Uzan-Narovlansky2022} or Bloch oscillations~\cite{Huber2014}. Theoretical studies have been mainly conducted in dielectrics, semiconductors and semi-metals like graphene~\cite{exp_HHG_graphene_Taucer, bilayerGraphene}. Metals, on the other hand, have remained largely unexplored, a consequence of their high reflectivity at infra-red wavelengths. Recent HHG experiments in thin films of a transition metal nitride (TiN), a transition metal dichalcogenide (NbSe$_2$) and a noble metal~\cite{Korobenko2021, Katayama2024, gholammirzaei2025highharmonicgenerationnoble, Takeda2HNbSe2} are starting to reveal that metallic crystals can also display highly non-linear optical responses (e.g., up to 25~eV in the case of silver) below the damage threshold, allowing to explore their ultrafast light-driven electron dynamics. Thus, it has become necessary to explore theoretically the characteristics of high-harmonic generation in the metallic regime, as compared to the more studied insulating and semiconducting regime.

In this context, the layered transition metal dichalcogenide 2H-NbSe$_2$ is a perfect candidate for this study; it can be directly compared to its well-studied semiconductor counterparts, e.g., 2H-MoS$_2$~\cite{Mette_Zuerch, Yoshikawa2019Interband, Liu2017, ChangLee2024}, and it has recently been shown to produce high harmonics \cite{Takeda2HNbSe2}. We point out that recent work using real-time time-dependent density functional theory has shown that the lack of inversion symmetry in both the 2H and CDW phases leads to even harmonic generation, as in its semiconducting analogues~\cite{RehnDanNbSe2}.

Here, we perform numerical HHG experiments on monolayer 2H-MoS$_2$ (semiconductor) and 2H-NbSe$_2$ (metal) by solving the equation of motion of the reduced density matrix equation in a real-space formulation~\cite{ATATA}. We account for electron-electron interaction effects using the time-dependent Hartree + static exchange (TD-HSEX) method (see Appendix~\ref{app:methods} and refs.~\cite{ATATA, Pereira_r0}), and consider 22 bands with spin-orbit coupling. Both materials share a hexagonal symmetry and have a strikingly similar band structure, with the key difference being the Fermi level. This allows to better pinpoint the differences between the metallic and semiconducting strong-field dynamics. In the following, we explore differences in the linear response, angle-resolved HHG spectrum, and population and recombination dynamics, highlighting the key characteristics of HHG in strongly-driven metallic crystals.

\section{Results}
\subsection{Linear conductivity and High Harmonic Generation \label{subsec:HHG}}

\begin{figure}
\begin{centering}
\includegraphics[width=0.7\textwidth]{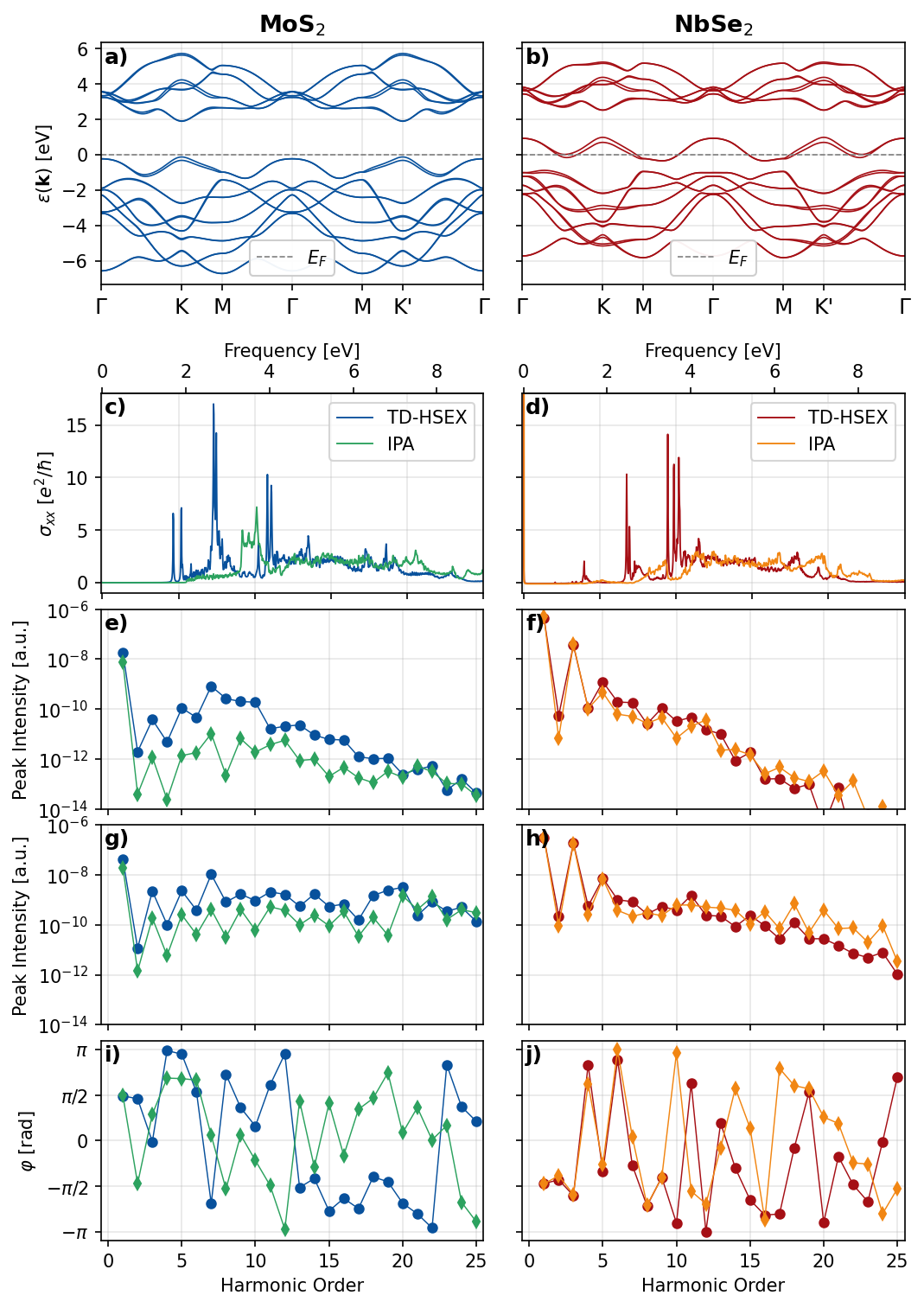}
\par\end{centering}
\caption{Electronic and optical properties of monolayer MoS$_{2}$ (left column) and NbSe$_{2}$ (right column). (a,b) DFT-calculated band structure used in the calculations, which includes spin-orbit coupling. (c,d) Real part of the linear optical conductivity $\sigma_{xx}(\omega)$. (e-h) HHG maximum yield for each harmonic for (e,f) I = 180~GW/cm$^2$ and (g,h) I = 500~GW/cm$^2$. (i,j) HHG phase at the maximum yield for I = 180~GW/cm$^2$. In panels c-h, we show results that include electron-electron interactions with TD-HSEX (blue for MoS$_{2}$, red for NbSe$_{2}$) and exclude them (IPA, teal for MoS$_{2}$, orange for NbSe$_{2}$). The HHG was obtained for a laser wavelength of $\lambda=3400$~nm and polarized along $\Gamma$--M. \label{fig: sigma_hhgs}}
\end{figure}

In Fig.~\ref{fig: sigma_hhgs}a,b we show the band structures of both semiconducting MoS$_2$ and metallic NbSe$_2$, calculated using a hybrid functional (details in Appendix~\ref{app:methods}). All of the calculations are performed for room temperature. We start by comparing their optical conductivity in Fig.~\ref{fig: sigma_hhgs}c,d. We either use an independent particle approximation (IPA) ignoring electron-electron interactions, or we take them into account by adopting the TD-HSEX method using a Rytova-Keldysh pseudopotential~\cite{Pereira_r0, ATATA,Edu_excitons}. The pseudopotential parameters, as well as details of the field-free density functional theory (DFT) calculations and the theoretical framework used for the temporal propagation are given in Appendix~\ref{app:methods}, and are further detailed in previous works~\cite{ATATA}. We note the absence in the literature of specific Rytova-Keldysh parameters to model electron-hole interaction effects in NbSe$_2$. Hence, we use the same Rytova-Keldysh parameters as for MoS$_2$ as an effective interaction model to capture qualitative TD-HSEX renormalization effects. We do not claim that they provide a quantitatively accurate screened interaction for the metallic monolayer, but rather use it to assess how electron-hole interaction effects qualitatively influence the HHG response in comparison with the semiconducting case, including trends that could be probed experimentally.

For MoS$_2$, the well-studied A and B excitons originating from the two spin-orbit-split valence bands appear at 1.7~eV and 1.9~eV, in good agreement with previous work~\cite{Pereira_r0}. In NbSe$_2$, we identify three sets of peaks that we associate to electron-hole interaction-enhanced transitions between the valence and metallic band, the metallic and the conduction band, and the valence and conduction band, as we will expand on later. In general, we observe a red-shift of the maxima of the optical conductivity due to electron-hole interactions, both in the semiconductor and the metal.

As reported in earlier works, interactions impact also the high-harmonic generation (HHG) spectrum of 2D semiconductors~\cite{Yoshikawa2019Interband, Edu_excitons, jensen_high-harmonic_2024}. To calculate the HHG spectrum, during the time evolution we include a site-distance-dependent dephasing term that damps coherences between orbitals that are far apart in the real-space representation~\cite{graham_real} (further details in Appendix~\ref{app:methods}). This suppresses long-range coherences and leads to smoother HHG spectra, bringing them closer to experimental observations by effectively mimicking macroscopic propagation and zero-point motion effects~\cite{graham_real,cardenas2025effectszeropointmotionhigh}. In addition, we use a dephasing time $T_2$ in the $k$-space representation to phenomenologically account for scattering processes. We use different values for each material, $T_2^{\text{(MoS$_2$)}}=100$~fs and $T_2^{\text{(NbSe$_2$)}}=10$~fs, reflecting the faster loss of coherence in the metallic system~\cite{Takeda2HNbSe2}. In Appendix~\ref{app:methods} we also show results using the same dephasing times for both materials, which do not change the main conclusions that we draw. Fig.~\ref{fig: sigma_hhgs}e,g display the HHG spectrum for MoS$_2$ using two different driving intensities polarized along the $\Gamma$-M direction, using both the IPA and TD-HSEX. We use a driving field with a Gaussian envelope of 50~fs full-width at half-maximum, wavelength of $\lambda=3400$~nm, and field strengths of $I=180$~GW/cm$^2$ (panel e) and $500$~GW/cm$^2$ (panel g). We note that the semiconductor has been shown to withstand higher fluences~\cite{Liu2017,Mette_Zuerch}, but they may be above the damage threshold of monolayer NbSe$_2$. Nonetheless, here we choose the same laser parameters in order to better compare the two materials, and note that using, e.g., a shorter driving pulse, should not affect the key results presented here. An enhanced yield in MoS$_2$ is observed when interactions are present, in line with previous works, especially for harmonics emitted within the first conduction band. For higher driver intensities and higher energies, multiple bands participate in the emission dynamics and interaction effects are weaker.

The HHG spectrum of the metallic system is shown in Fig.~\ref{fig: sigma_hhgs}f,h for a laser polarized along $\Gamma$-M. The effect of interactions is weaker than in the semiconducting case. While this is partly influenced by the shorter dephasing time $T_2$ of the metal, we find that the effect of interactions is lower in the metal even for the same dephasing as the semiconductor (see Appendix~\ref{app:methods}). The lowest odd-order harmonics show a considerably stronger yield than in the semiconductor, which is associated to strong intraband currents from the partially occupied metallic bands and their nonparabolic dispersion near the Fermi level. Increasing the intensity from 180~GW/cm$^2$ to 500~GW/cm$^2$ mostly elongates the plateau but does not change the relative effect of electron-hole interactions. 

Fig.~\ref{fig: sigma_hhgs}i,j shows that the electron-hole interactions not only affect the HHG yield, but also the phase of the harmonics. The phase change induced by interactions reaches up to $\sim~\pi$ for some harmonics, both in the semiconducting and metallic systems.

\begin{figure}[H]
\begin{centering}
\includegraphics[width=0.7\textwidth]{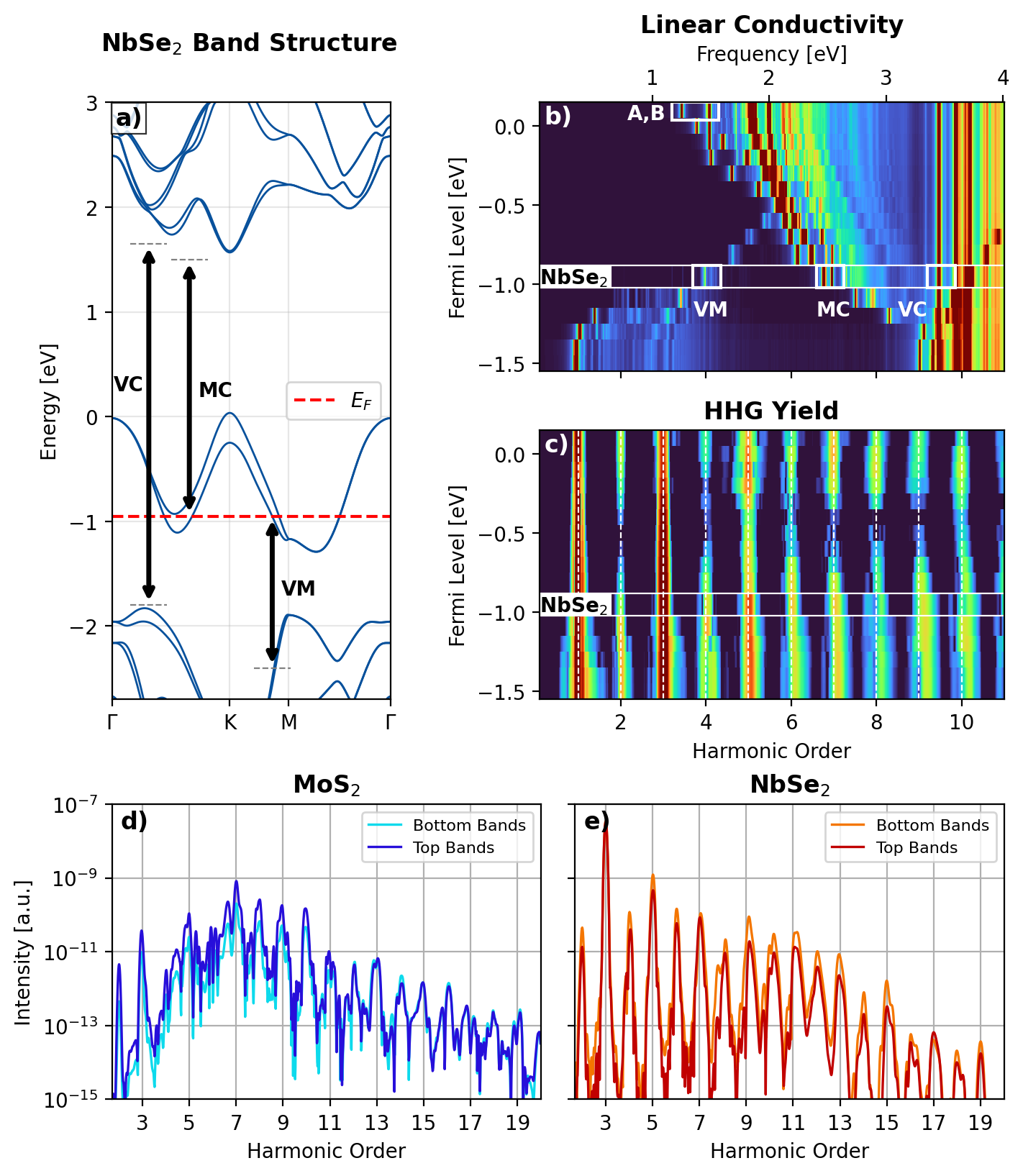}
\par\end{centering}
\caption{(a) Main linear optical conductivity peaks overlaid on the band structure of NbSe$_2$. (b) Real part of the linear optical conductivity $\Re[\sigma_{xx}(\omega)]$. (c) High harmonic spectrum as a function of harmonic order and Fermi level $E_F$. Four prominent resonances are labeled in the conductivity, corresponding to transitions between valence-to-metallic band (VM), metallic-to-conduction band (MC) and valence-to-conduction band (VC) in NbSe$_2$, and excitons A and B when $E_F\geq$0.0~eV. (d,e) HHG of MoS$_2$ (d) and NbSe$_2$ (e) calculated by summing the currents from the two spin-splitted bands that are below the empty bands (i.e., the highest valence bands in MoS$_2$ and metallic bands in NbSe$_2$) with either the bands below (bottom bands, light blue) or above (top bands, dark blue). HHG yields are calculated with a laser polarized in $\Gamma$--M.}
\label{fig: grow_EF}
\end{figure}

To understand the effect of interactions in the metal, the conductivity peaks and their impact in the harmonic yield are explored in Fig.~\ref{fig: grow_EF}~a-c by varying the Fermi level of NbSe$_2$. We do so by applying a rigid shift, without changing any other parameter of the system. As mentioned earlier, we identify three sets of peaks in the optical conductivity (Fig.~\ref{fig: grow_EF}a), associated to transitions between the valence and metallic band (VM), the metallic and the conduction band (MC), and the valence and conduction band (VC). This association is clear by looking at the plot of the optical conductivity as a function of the Fermi level (Fig.~\ref{fig: grow_EF}b): the VM and MC peaks shift toward higher and lower energies, respectively, as the Fermi level is increased, while the VC peaks remain unchanged.

In Fig.~\ref{fig: grow_EF}c we now plot the low-energy HHG spectrum as a function of the Fermi level shift. Interestingly, we find a clear blue shift in the central energy of the harmonics with $\hbar\omega > 2.2$~eV as we decrease the Fermi level. Note that for a Fermi level of $E_F=-1.5$~eV, the system is once again a semiconductor, with the metallic band now completely empty. The reason why the central harmonic energy appears shifted in this case, as well as in the metallic case, is that the minimum direct band gap occurs at several $k$-points that are not related by fundamental symmetries (e.g., inversion or time-reversal). In that case, different regions of the Brillouin zone will be excited with distinct excitation and saturation dynamics, leading to a time-dependent imbalance in their contributions (e.g., due to Pauli blocking and population buildup). As a result, the emission is biased toward the rising edge of the pulse (see Appendix~\ref{app:blueshift}), which leads to a blue shift of the harmonic peak~\cite{PhysRevLett.113.193901}. In contrast, when the band structure valleys are symmetry-related, their dynamics remain balanced and the emission remains centered around the pulse maximum, so that no shift is observed. Besides this, we also observe a minimum of the high harmonic emission when the metallic band is approximately half-full ($E_F \approx -0.5$~eV). This corresponds to the case where the direct gap between the filled valence bands and the partially-filled metallic band is the same as that between the empty conduction and metallic band (see Fig.~\ref{fig: grow_EF}b).

One relevant question is if the high harmonics observed in the metallic system are generated by intraband currents from the two spin-splitted metallic bands or coherences between those two bands and the conduction or lower-lying valence bands. To address this, in Fig.~\ref{fig: grow_EF}~d,e we calculate the HHG spectrum of the semiconductor/metal from the sum of the currents generated by the valence/metallic spin-splitted bands and (i) the set of 8 conduction bands (top bands), (ii) the set of 12 low-lying valence bands (bottom bands), following an analogous current separation as in \cite{jimenez-galan_orbital_2023}. The fact that the top bands dominate in the semiconductor for harmonics below H12 (Fig.~\ref{fig: grow_EF}d) tells us that they are contributed mostly by dynamics involving higher conduction bands, either via intraband motion or interband coherences with the two spin-splitted valence bands. Harmonics above H12, on the other hand, show very similar contribution from the top and bottom bands, indicating that they originate from coherences involving almost equally dynamics from low-lying valence and higher conduction bands. In contrast, in the metallic system (Fig.~\ref{fig: grow_EF}e), there is a consistent dominant contribution of low-lying valence bands throughout the HHG spectrum. The fact that the top and bottom contributions are not equal indicates that the HHG signal is not dominated solely by intraband currents within the metallic bands. Since these bands are included in the top and bottom partitions, a purely intraband response would contribute similarly to both. The observed imbalance therefore points to a significant role of interband coherences involving low-lying valence bands.

\begin{figure}
\begin{centering}
\includegraphics[width=.8\textwidth]{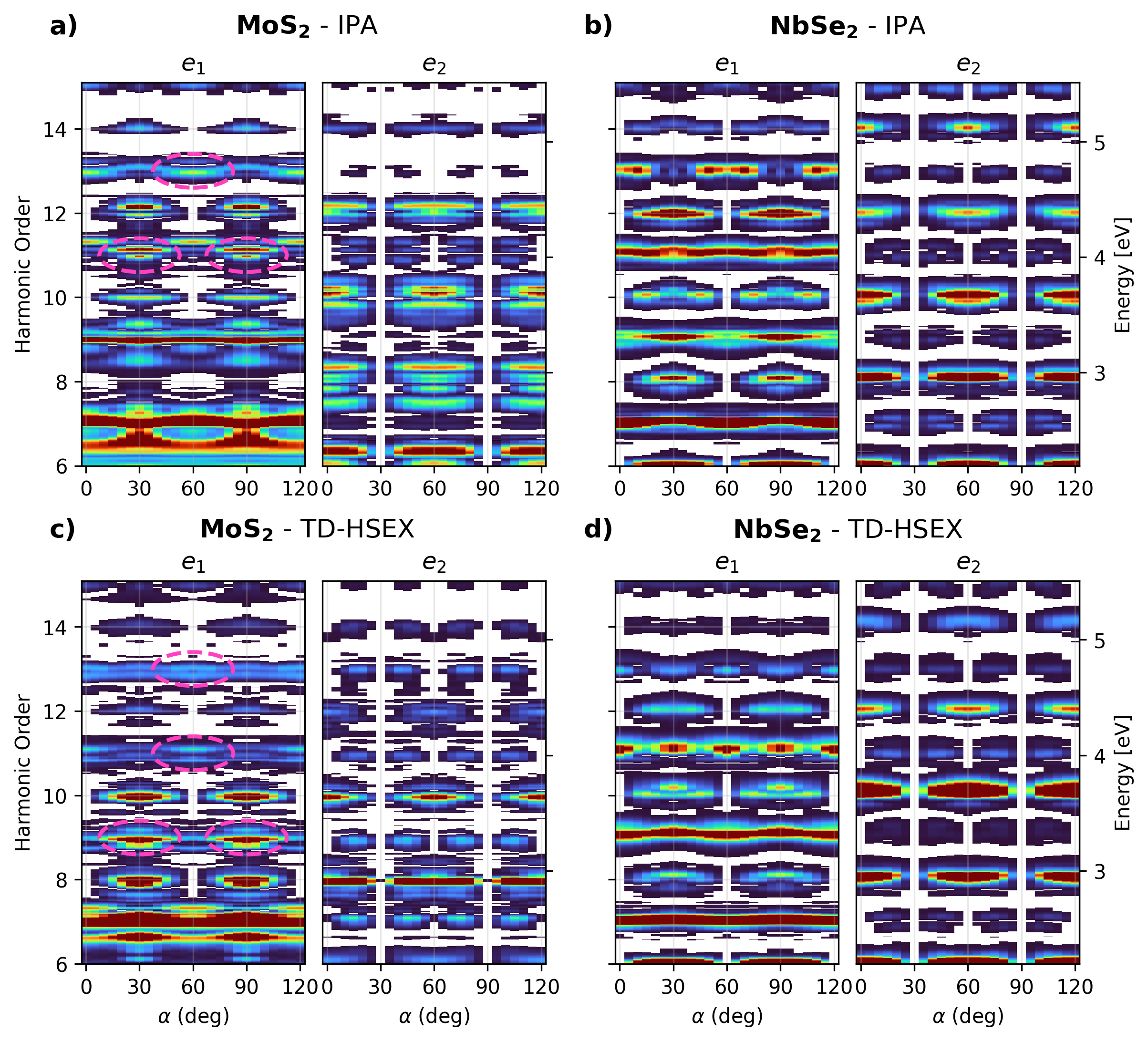}
\end{centering}
\caption{Angle-resolved high-harmonic spectra of (a,c) MoS$_{2}$ and (b,d) NbSe$_2$ along the parallel ($e_1$) and perpendicular ($e_2$) directions relative to the laser field polarization. Panels (a,b) correspond to the IPA wheras panels (c,d) correspond to TD-HSEX. Directions $\Gamma$--K and $\Gamma$--M correspond to $\alpha=0^{\circ}$ and $30^{\circ}$, respectively. The pink circles indicate the change in angular pattern (see text for details).  \label{fig: colorplots}}
\end{figure}

We now turn to the full angle-resolved HHG spectra, shown in Fig.~\ref{fig: colorplots}. Both the semiconductor and metal reflect the dynamical symmetry of the lattice~\cite{Liu2017, Mette_Zuerch}: along $\Gamma$--K ($\alpha=0^\circ$) only odd (even) harmonics are emitted in the parallel (perpendicular) direction relative to the driving pulse, while along $\Gamma$--M ($\alpha=30^{\circ}$) all harmonics are emitted parallel to the driver and no harmonics are emitted in the perpendicular direction. Yue et al. recently reported that the angle-resolved HHG spectrum of MoS$_2$ revealed multi-band effects, which were responsible for an angular shift of the HHG maxima from $\Gamma$-M to $\Gamma$-K above 3.5~eV~\cite{Mette_Zuerch}. We find that this feature  also has a fingerprint of excitonic effects. As Fig.~\ref{fig: colorplots}a shows, in absence of interactions such angular shift occurs for higher energies, between H11 and H13. When interactions are included (Fig.~\ref{fig: colorplots}b), the angular shift occurs between H9 and H11, as observed experimentally (cf. Fig.~1 of~\cite{Mette_Zuerch}).

For NbSe$_{2}$, shown in Fig.~\ref{fig: colorplots}b,d, we also observe similar changes in the angular patterns along the laser polarization direction. In absence of interactions, the $\Gamma$-M ($\alpha=30^\circ$) direction shows a maxima for H9 and minima for H11 and H13. When interactions are included, both $\Gamma$-K and $\Gamma$-M directions show maxima for H9-H13 along the laser polarization direction. This contrasts to the semiconductor, where we observe always a maxima along $\Gamma$-K when $\Gamma$-M is minimum, and viceversa. Along the perpendicular direction to the laser polarization, both semiconductor and metal share a similar angular spectrum.

\subsection{Electronic population dynamics \label{subsec:dynamics}}

\begin{figure}
\begin{centering}
\includegraphics[width=.8\textwidth]{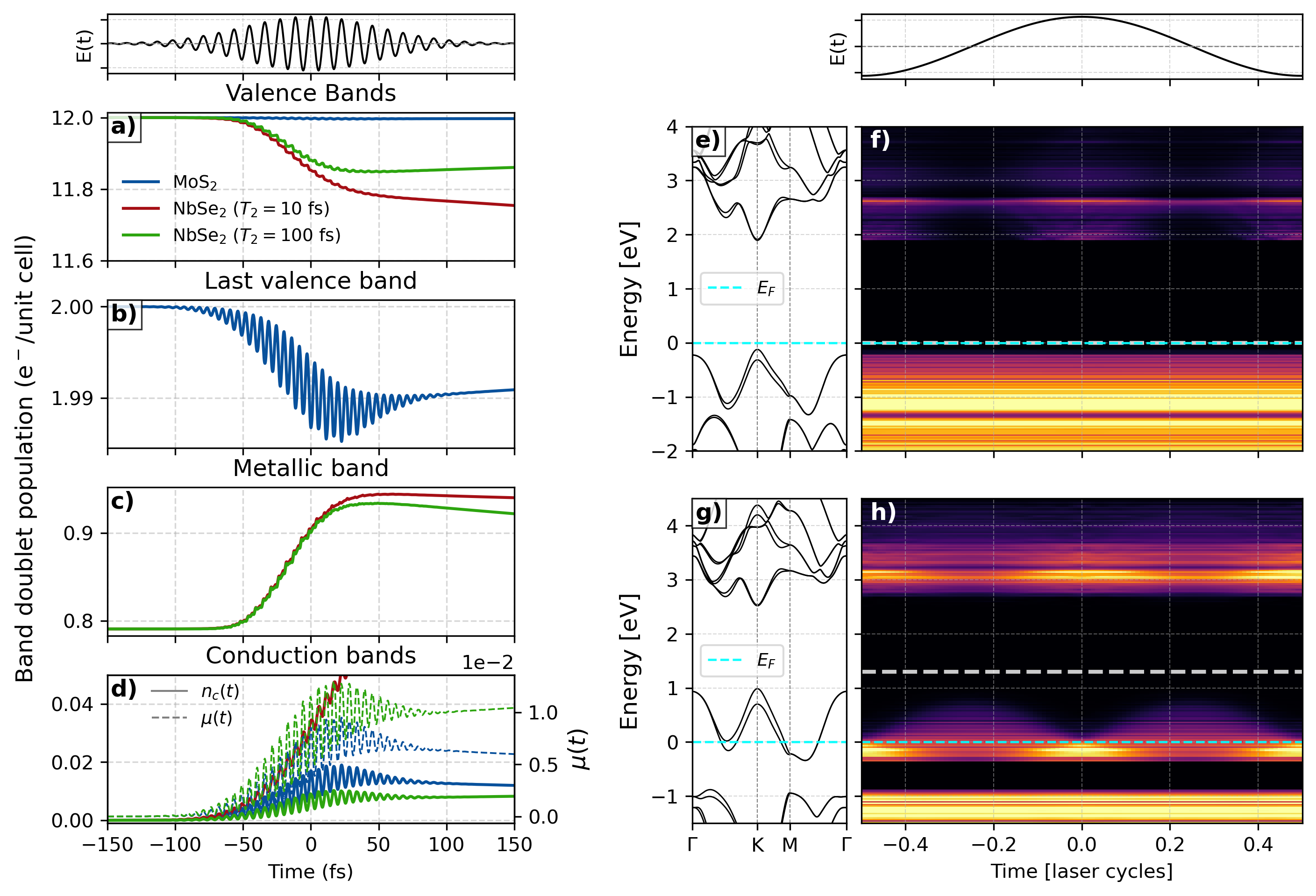}
\end{centering}
\caption{ (a--d)~Time-resolved populations in MoS$_2$ and NbSe$_2$ (in electrons per unit cell) of: (a)~low-lying valence bands (i.e., excluding highest spin-splitted valence bands in MoS$_2$ and metallic bands in NbSe$_2$), (b)~highest spin-splitted valence bands in MoS$_2$, (c)~spin-splitted metallic bands in NbSe$_2$ only, (d - left axis, solid lines)~conduction bands. (d - right axis, dashed lines)~Carrier injection efficiency $\mu(t) = n_c(t) / n_0$, where $n_c(t)$ is the time-dependent population of the conduction bands and $n_0$ is the initial population of the last valence (metallic) bands of MoS$_2$ (NbSe$_2$) with $T_2=100$~fs. (e--h)~Electronic band structures along the $\Gamma$--K--M--$\Gamma$ path and the corresponding energy-resolved time-dependent populations weighted by the density of states over one laser cycle for (e, f) MoS$_2$ and (g, h) NbSe$_2$. The dashed cyan lines mark the Fermi level. The same color scale is used for both materials, but we have used a different color-scale normalization for the energy region of the conduction bands, delimited by the white dashed line.  The laser field has an intensity of I = 180 GW/cm$^2$ and is polarized in $\Gamma$--M.} 
\label{fig: popus}
\end{figure}

We focus now on the electron population dynamics. Fig.~\ref{fig: popus}a-d shows the energy-resolved time-dependent populations for both MoS$_2$ and NbSe$_2$. As we discussed earlier, the low-lying valence bands are more depleted in the metallic system, concomitant with the increased population of the metallic band. We also observe that the number of excited electrons to the conduction bands of the metal is larger than for the semiconductor. This is in part an effect of the shorter decoherence time $T_2$ of the metal, which transforms virtual excitations into real excitations by impeding the virtual excited electrons from returning to the ground state~\cite{Boroumand_2025}. While the semiconductor shows a higher number of excited electrons $n_c$ when both dephasing times are chosen equal (green curves in Fig.~\ref{fig: popus}a-d), one should note that the total number of available electrons for excitation is considerably smaller in the metal. A better comparison of the carrier injection efficiency in both materials is given by the dashed curves in Fig.~\ref{fig: popus}d, which gives the number of excited electrons normalized by the initially available electrons $n_0$ in the dominant source bands, i.e., the two spin-splitted valence bands in MoS$_2$ ($n_0=2$~e$^-$/unit cell) and the two spin-splitted metallic bands in NbSe$_2$ ($n_0=0.79$~e$^-$/unit cell), $\mu(t) = n_c(t)/n_0$. In this case, we see that, even for equal $T_2$, NbSe$_2$ has a higher carrier injection efficiency, though it produces fewer total excited electrons. This might come as a surprise given its larger energy gap between the Fermi level and the conduction band minimum.

To understand this, we note that in the strong-field regime interband excitation occurs predominantly via field-induced tunneling and is therefore exponentially sensitive to the electric field amplitude $\left| E(t) \right|$ and the instantaneous minimum energy separation between the participating bands. Assuming for simplicity that the bands are not largely modified by the field, this effective minimum energy gap during the light-matter interaction will be $\Delta \varepsilon (\mathbf{k}(t)) = \varepsilon_c (\mathbf{k}(t)) - \varepsilon_v (\mathbf{k}(t))$, where $\varepsilon_{c/v}$ are the energies of the first conduction and last valence bands at $\mathbf{k}(t)$, respectively, and $\mathbf{k}(t)=\mathbf{k}_0+\mathbf{A}(t)$ is the streaked crystal momentum. The crystal momentum $\mathbf{k}_0$ is that which corresponds to the minimum band gap in equilibrium, and $\mathbf{A}(t)$ is the time-dependent vector potential. For a semiconductor, $\Delta \varepsilon(\mathbf{k}(t)) \ge \Delta \varepsilon(\mathbf{k}_0)$, since the valence band is initially filled. Moreover, for a linearly-polarized field, $\mathbf{A}(t_i)=0$ at the maximum of the electric field $t_i$. Thus, we expect electron injection to occur at the maximum of the electric field and at $\mathbf{k}_0$, which corresponds to $\pm K$ in the case of MoS$_2$. In Fig.~\ref{fig: popus}f, we plot the energy-resolved time-dependent populations of MoS$_2$ weighted by the static density of states. We observe half-cycle-periodic population injection at the maximum of the field, and at the energy corresponding to the equilibrium minimum band gap $\pm K$. This situation is different for the metal. The time-dependent effective band gap can now be smaller than the equilibrium band gap, $\Delta \varepsilon(\mathbf{k}(t)) \le \Delta \varepsilon(\mathbf{k}_0)$, since it has a partially-filled band. The energy-resolved time-dependent populations of the metal show no conduction band population at the energy of the $\pm K$ points, but instead display a much higher population at an energy of $\sim 3$~eV as compared to the semiconductor, which hosts a high density of states. Note also that the energy difference between the maximum of the metallic band and the minimum of the conduction band is smaller than the band gap of MoS$_2$. Hence, the effective conduction band gap during the laser-matter interaction in the metal can be at times smaller than in the semiconductor, which can explain the higher injection efficiency. 

\begin{figure}
\begin{centering}
\includegraphics[width=0.5\textwidth]{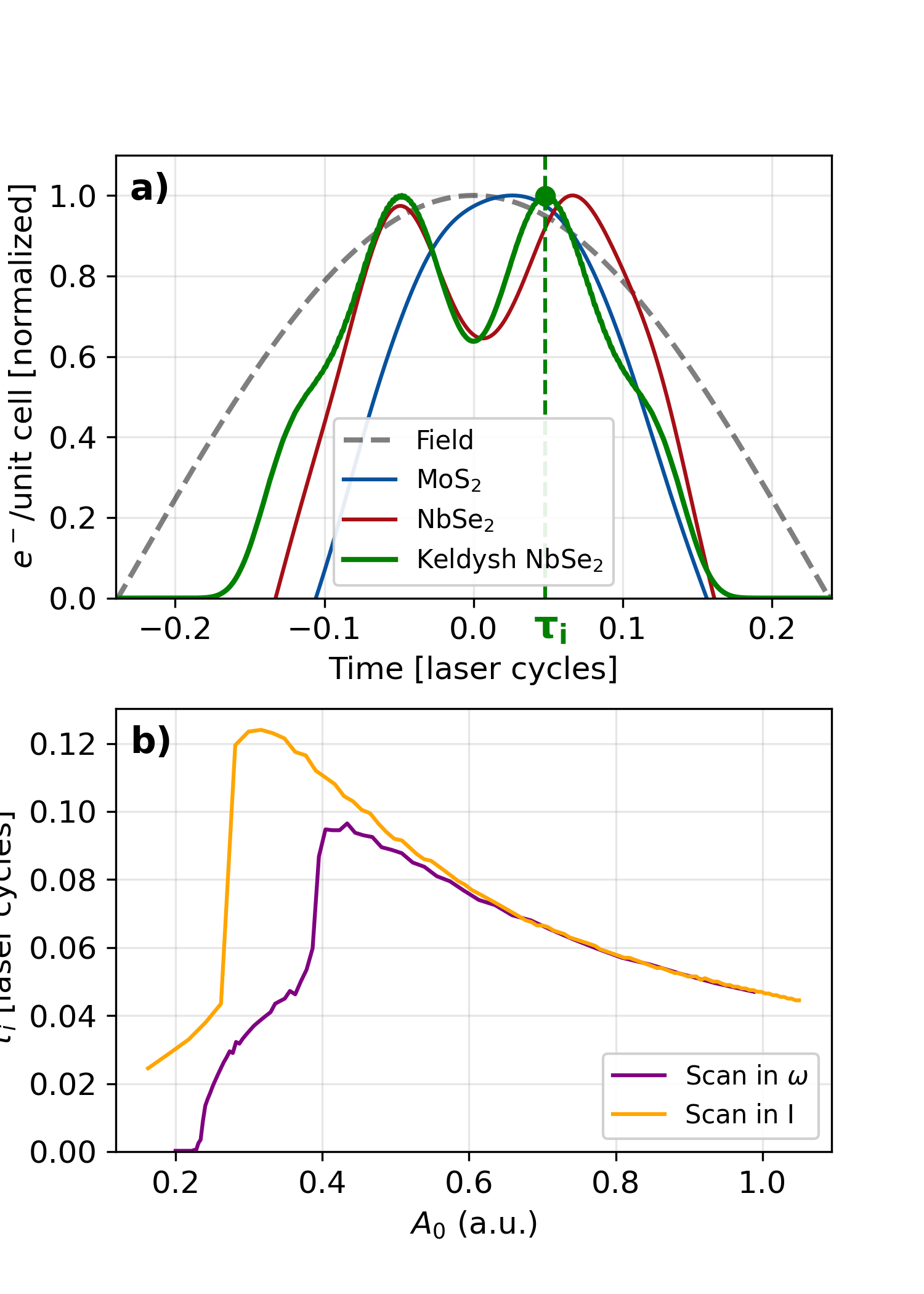}
\par\end{centering}
\caption{(a) Normalized oscillatory component of the spin-splitted conduction band population (see text for details) of MoS$_2$ (blue) and NbSe$_2$ (red) for a single-cycle pulse polarized along the $\Gamma$-M direction, shown as the grey dashed line. The Keldysh tunneling ionization rate for last valence and first conduction band of NbSe$_2$ is shown in green, with $\tau_i$ being the time of maximum ionization. (b) Time of maximum ionization $\tau_i$ as a function of the vector potential magnitude $A_0$, calculated with Eq.~\ref{eq:tunnel_ionization} by performing a scan in frequency $\omega$ with constant intensity $I$ (purple), and for a scan in intensity $I$ with constant $\omega$ (orange). For simplicity, we have used a constant reduced interband effective mass, $m(k(t))^* = 1$ (see Eq.~\ref{eq:tunnel_ionization}). The laser is a single cycle gaussian with I~=~500~GW/cm$^2$.} \label{fig: ion_rate}
\end{figure}

In the metal, the time and crystal momentum of maximum electron injection to empty bands depend heavily on the pulse characteristics, in particular on the balance between a high electric field amplitude and a low effective band gap. Fig.~\ref{fig: ion_rate}a shows the oscillatory component of the conduction-band doublet population in NbSe$_2$ and MoS$_2$, extracted by removing the slowly varying contribution via one-period centered temporal averaging, i.e. $n_{osc}(t)=n(t)-\left<n\right>_T(t)$. While in MoS$_2$ the maximum tunneling probability happens close to the electric field maximum, in NbSe$_2$ it is clearly shifted, displaying two almost symmetrical peaks with respect to the field maximum. For simplicity, we have considered a single-cycle pulse in this case. The small differences between the two peaks are due to finite-pulse effects, transient populations and dephasing effects. This two-peak structure is qualitatively reproduced by the instantaneous tunneling ionization rate in the low-frequency and strong-field regime~\cite{Keldysh:1965ojf},
\begin{equation}\label{eq:tunnel_ionization}
    \omega_0(t) \propto \exp \left( -\frac{4\,\sqrt{2m(\mathbf{k}(t))^*}\,\Delta \varepsilon(\mathbf{k}(t))^{3/2}}{3\,{|E(t)|}}\right),
\end{equation}
where $|E(t)|$ is the electric field strength, $m(\mathbf{k}(t))^*$ is the reduced interband effective mass, and $\Delta\varepsilon(\mathbf{k}(t))$ is the instantaneous band separation evaluated here between the metallic band and the lowest unoccupied conduction band in NbSe$_2$. Hence, in the metal, both the maximum ionization time and crystal momenta can be controlled by the shape of the electric field. To compare with the NbSe$_2$ result in Fig.~\ref{fig: ion_rate}a, we evaluate $\Delta \varepsilon(\mathbf{k}(t))$ from the metallic band and the lowest unoccupied conduction band, and integrate Eq.~\ref{eq:tunnel_ionization} over the equilibrium initially populated crystal momenta $\mathbf{k}_0$ in the Brillouin zone. In Fig.~\ref{fig: ion_rate}~b we calculate this $\mathbf{k}_0$-integrated tunneling ionization rate $\omega_0(t)$ for a scan of intensity and frequency of the laser pulse, showing how the injection time can be controlled by these laser parameters in NbSe$_2$. The maximum delay time $\tau_i$ between the time of maximum injection and the field peak can reach up to 0.12 laser cycles.

\begin{figure}
\begin{centering}
\includegraphics[width=0.8\textwidth]{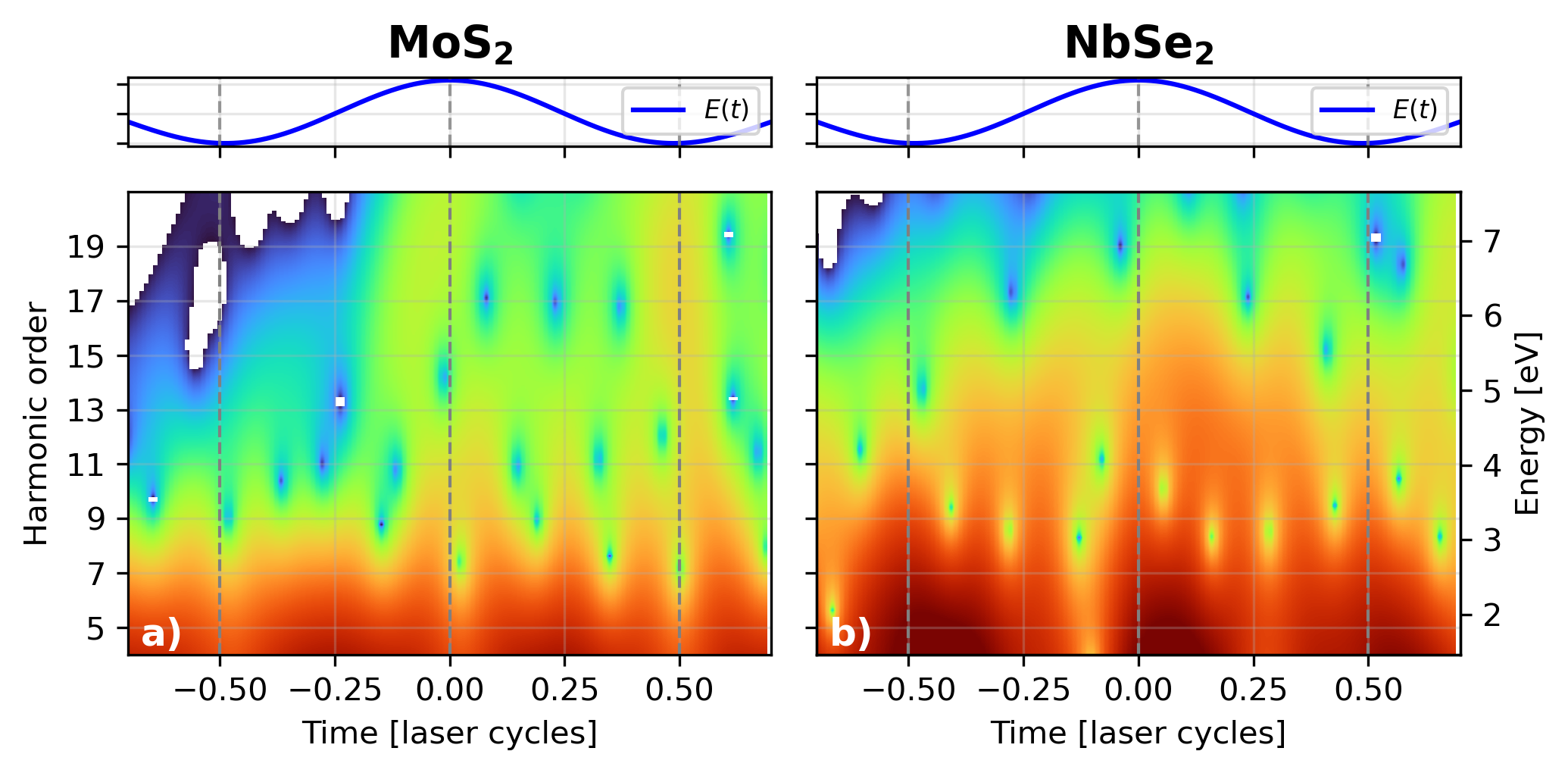}
\par\end{centering}
\caption{Time-frequency analysis of high-harmonic generation (HHG) in (a) MoS$_2$  and (b) NbSe$_2$ via Gabor transform with a Gaussian window with variance $\alpha=(3\omega)^{-1}$. Top panels show the driving $E(t)$, which is chosen as a single-cycle pulse in this case. Color maps display the logarithmic intensity of emitted harmonics as a function of time (in laser cycles) and harmonic order, with the corresponding photon energy shown on the right axis. Grey dashed lines indicate the field extrema. The laser is polarized along $\Gamma$--M.} 
\label{fig: Gabor}
\end{figure}

Finally, in Fig.~\ref{fig: Gabor} we compare the recombination dynamics for a single-cycle pulse through time-frequency Gabor maps, which reveal a clear difference in the timing of maximum harmonic emission. In the semiconductor case, and similarly to atomic systems, harmonics below H13 are not emitted at the field maxima. These harmonics originate predominantly from interband recombination between the highest valence and lowest conduction bands~\cite{VampaTutorial}. In contrast, in the metallic case, the field peaks do support strong harmonic emission, reflecting the intraband acceleration of carriers near the Fermi level in strongly non-parabolic bands. At the same time, significant emission at other times, particularly for higher harmonics, indicates the presence of both intraband and interband contributions.

\section{Conclusion}
In conclusion, we have used accurate multi-band simulations to explore the fundamental differences in the highly non-linear optical response of a realistic 2D semiconductor and metal, including time-dependent electron-hole interaction effects. We have first observed that the linear optical response of the metallic system shows a series of peaks when correlation effects are considered. These are associated to three types of transitions between the bands: valence-to-metallic, valence-to-conduction and metallic-to-conduction. As reported in previous works, in the semiconductor such correlation effects lead to an enhancement of near-band-gap harmonics due to the bound excitons. We do not observe such a strong enhancement for the metallic case, but we do observe a relevant harmonic phase variation. We also observed the impact of electron-hole correlation effects in the angular-resolved HHG, which e.g., shifts towards lower energies the change in the angular pattern of MoS$_2$, bringing it to agreement with recent measurements~\cite{Mette_Zuerch}. In contrast to the semiconductor, we saw that low-lying valence bands in the metal play a key role in the harmonic generation consistent with the smaller energy separation; a result similar to that observed in silver~\cite{gholammirzaei2025highharmonicgenerationnoble}. The population dynamics driven from the metallic band to the first empty conduction band were also shown to differ strongly from those between the valence and conduction bands in the semiconductor. While in the latter these population dynamics are triggered predominantly near the maxima of the electric field, in the metal their timing depends strongly on the band dispersion and the parameters of the electric field. In our regime, we have shown that they can be qualitatively captured by Keldysh tunneling ionization rate. Furthermore, we have shown that the injection efficiency to empty bands is higher in NbSe$_2$, despite the total number of excited electrons being smaller. Finally, we have seen that strong harmonic emission in the metal can occur at the maxima of the electric field, in contrast to the semiconducting case, reflecting strong intraband contributions even for harmonics well above the metallic to conduction band energy gap.

\begin{acknowledgments}
This work was supported through the Talento Comunidad de Madrid Fellowship 2022-T1/IND24102, the Spanish Ministry of Science, Innovation
and Universities through grant reference PID2023-146676NA-I00, and CNS2025-166331. R.E.F. Silva acknowledges support from the Spanish Ministry of Science, Innovation
and Universities through grants RYC2022-035373-I and CNS2024-154463.
\end{acknowledgments}

\section*{Data Availability Statement}
The data that support the findings of this study are available from the corresponding author upon reasonable request.

\appendix

\section{Methods \label{app:methods}}
The field-free Hamiltonian of both materials is computed using the density functional theory (DFT) code QuantumEspresso~\cite{Quantum_Espresso} with a Heyd-Scuseria-Ernzerhof (HSE) hybrid exchange-correlation functional in a Monkhorst-Pack grid of 16x16x2. We then map the $k$-space Hamiltonian onto a tight binding Hamiltonian written in the basis of maximally-localized Wannier functions using the Wannier90 code~\cite{WANNIER}. We reduce the size of each crystal Hamiltonian to a total of 22 bands by projecting onto the atomic-like $s$ and $p$ orbitals of the chalcogen atom and $d$ orbitals of the transition metal, including spin-orbit interaction. The quasiparticle band gap of the semiconductor is 2.02 eV. 

The HHG spectrum is computed from the Fourier transform of the total current induced in the crystal by the strong laser field~\cite{ATATA},
\begin{equation}
\mathbf{J}(t)=\left|e\right|\sum_{\boldsymbol{R}}\sum_{nm}\mathbf{v}_{nm}(\boldsymbol{R})\rho_{mn}(-\boldsymbol{R},t),
\end{equation}
where $\mathbf{v}(\mathbf{R}) = i \left( \mathbf{R} \hat{H}_0(\mathbf{R}) - \sum_\mathbf{R'}[\hat{\mathbf{r}}(\mathbf{R'}),\hat{H}_0 (\mathbf{R}-\mathbf{R}')]\right)$ is the velocity operator written in real space.

In order to obtain the current, we use the real-time, real-space reduced density matrix approach described in \cite{ATATA}. Briefly, we solve the equation of motion using the Hartree+screened static exchange approximation (HSEX) to account for electron-electron interaction,
\begin{align}
    \partial_t \rho_{ij}(t) =& -i \left[ \hat{H}_0 + \hat{\Sigma}^{HSEX}[\hat{\rho}(t)-\hat{\rho}^{(0)}]+ \mathbf{E}(t)\cdot\hat{\mathbf{r}},\hat{\rho}(t) \right]_{ij} \notag \\
    &+i\mathcal{L}_{incoh}\left[\rho_{ij}(t) - \rho_{ij}^{(0)} \right]
\end{align}
where $\hat{H}_0$ and $\hat{\mathbf{r}}$ are the field-free Hamiltonian and position operators obtained from DFT and Wannier90 simulations and $\mathbf{E}(t)$ is the electric field. 

The term $\hat{\Sigma}^{HSEX}$ contains the Hartree and Fock self energies in the Hartree + static screened-exchange (H-SEX) approximation, with $\hat{\rho}^{(0)}$ being the density matrix of the $H_0$ Hamiltonian in thermodynamic equilibrium. To simulate the electron-hole interaction, we employ a Rytova-Keldysh potential for the static screened-exchange term,
\begin{equation}
    W^{RK}(r)=\frac{\pi}{(\epsilon_{1}+\epsilon_{2})r_{0}}\left[H_{0}\left(\frac{r}{r_{0}}\right)-Y_{0}\left(\frac{r}{r_{0}}\right)\right],
\end{equation}
where $\epsilon_1$ and $\epsilon_2$ are the dielectric constants of the top and bottom medium, $r_0$ is the screening length and $H_0$, $Y_0$ are the zero-order Struve and Neumann special functions, respectively. We use the dielectric constants of air ($\epsilon_1=1$) and silica ($\epsilon_2=4$) with a screening length $r_{0}$ = 13.55 {\AA}, known to appropriately reproduce the optical conductivity spectra for MoS$_{2}$ with a silica substrate~\cite{RubioScreening,Pereira_r0}. For NbSe$_2$, the same Rytova-Keldysh parameters are used as an effective interaction model, noting that we do not claim they provide a quantitatively accurate description of electron-hole interaction for the metallic monolayer, but rather a qualitatively trend of the interaction effects.

\begin{figure}
\begin{centering}
\includegraphics[width=0.8\textwidth]{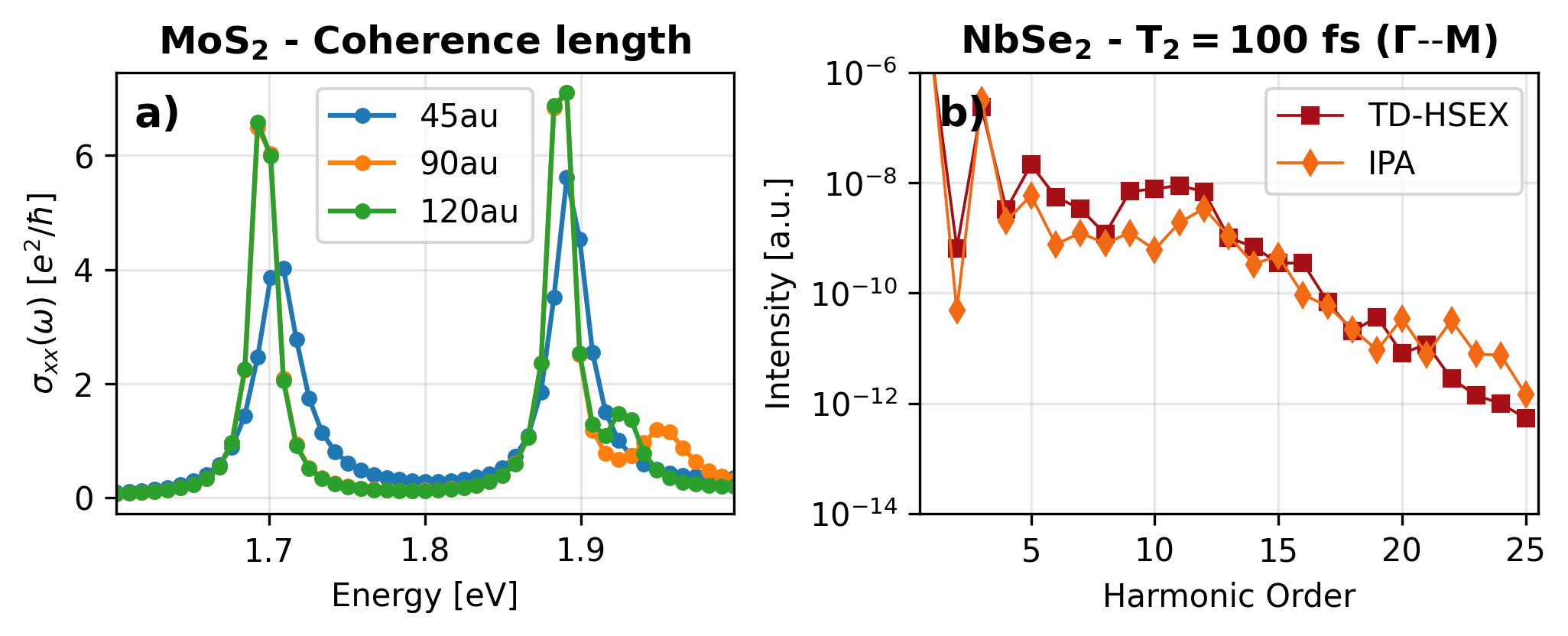}
\par\end{centering}
\caption{(a) Linear conductivity $\sigma_{xx}(\omega)$ for MoS$_2$ using a coherence length of 45 a.u. (blue), 90 a.u. (orange), and 120 a.u. (green). (b) HHG spectra of NbSe$_2$ for $T_2=100$ fs and a laser polarized along the $\Gamma$--M direction. Red curve includes $e^-e^-$ interactions (TD-HSEX) and orange curve does not (IPA). \label{fig:deph}}
\end{figure}

Finally, the incoherent term
\begin{equation}
\mathcal{L}_{incoh} = \mathcal{L}_{r} + \mathcal{L}_{\mathcal{D}} + \mathcal{L}_{rs}
\label{eq: incoh_eq}
\end{equation} 
includes the relaxation time $T_1$, dephasing time $T_2$, and a distance-dependent dephasing, respectively. Relaxation is assumed slow enough as to not affect the HHG spectrum and is set to $T_1^{\text{(NbSe}_2)}=T_1^{\text{(MoS}_2)} = 500$~fs. Dephasing times are chosen differently in the semiconductor and in the metal, reflecting the expected faster loss of coherence in the latter~\cite{Takeda2HNbSe2}: $T_2^{\text{(NbSe}_2)}=10$~fs, $T_2^{\text{(MoS}_2)} = 100$~fs. It is known that, while physically justified, using dephasing times longer than a few periods of the field leads to a noisy HHG spectrum~\cite{VampaTutorial,FlossPropagate,graham_real,cardenas2025effectszeropointmotionhigh}. This is because experimental HHG contains additional sources of decoherence that are not captured by the microscopic, single-particle reduced density matrix equation. To circumvent this, we additionally include a dephasing term $\mathcal{L}_{rs}$ that mimics macroscopic propagation and atomic jitter effects~\cite{graham_real,ATATA,cardenas2025effectszeropointmotionhigh}. This term removes density matrix coherences which are long range in a real-space formulation, 
\begin{equation}
    \mathcal{L}_{rs} [\rho-\rho^{(0)}](\mathbf{R}) = - \gamma(d(\mathbf{R})) \odot (\rho(\mathbf{R})-\rho^{(0)}(\mathbf{R})),
\end{equation}
where $\gamma$ is a parameter controlling the strength of the dephasing that is dependent on the distance between the orbitals $d(\mathbf{R})$, and $\odot$ denotes the Hadamard product. Effectively, $\gamma$ keeps the full coherence between orbitals that are separated less than coherence length value $\ell$, and suppresses those that are separated by more than $\ell$ quadratically with distance.  In order to choose $\gamma$, we make sure that the wavefunctions of both the A and B excitonic peaks of MoS$_2$ are well contained within our propagation box.  Fig.~\ref{fig:deph}a shows that a coherence length $\ell=90$~a.u. is enough for this purpose. Such choice of $\gamma$ does not affect NbSe$_2$, whose dephasing dynamics is governed by $T_2^{\text{(NbSe}_2)}$, which was chosen one order of magnitude smaller than that of MoS$_2$ (see above). Fig.~\ref{fig:deph}b shows the HHG spectra of NbSe$_2$ for the same conditions as Fig.~\ref{fig: sigma_hhgs}f, but using the same dephasing time $T_2=100$~fs as in the semiconductor. As expected, this leads to a slightly enhanced effect of electron-hole interactions, albeit only in the vicinity of H10.

\section{Harmonic blue-shift in NbSe$_2$, \label{app:blueshift}}
\begin{figure}
\begin{centering}
\includegraphics[width=0.7\textwidth]{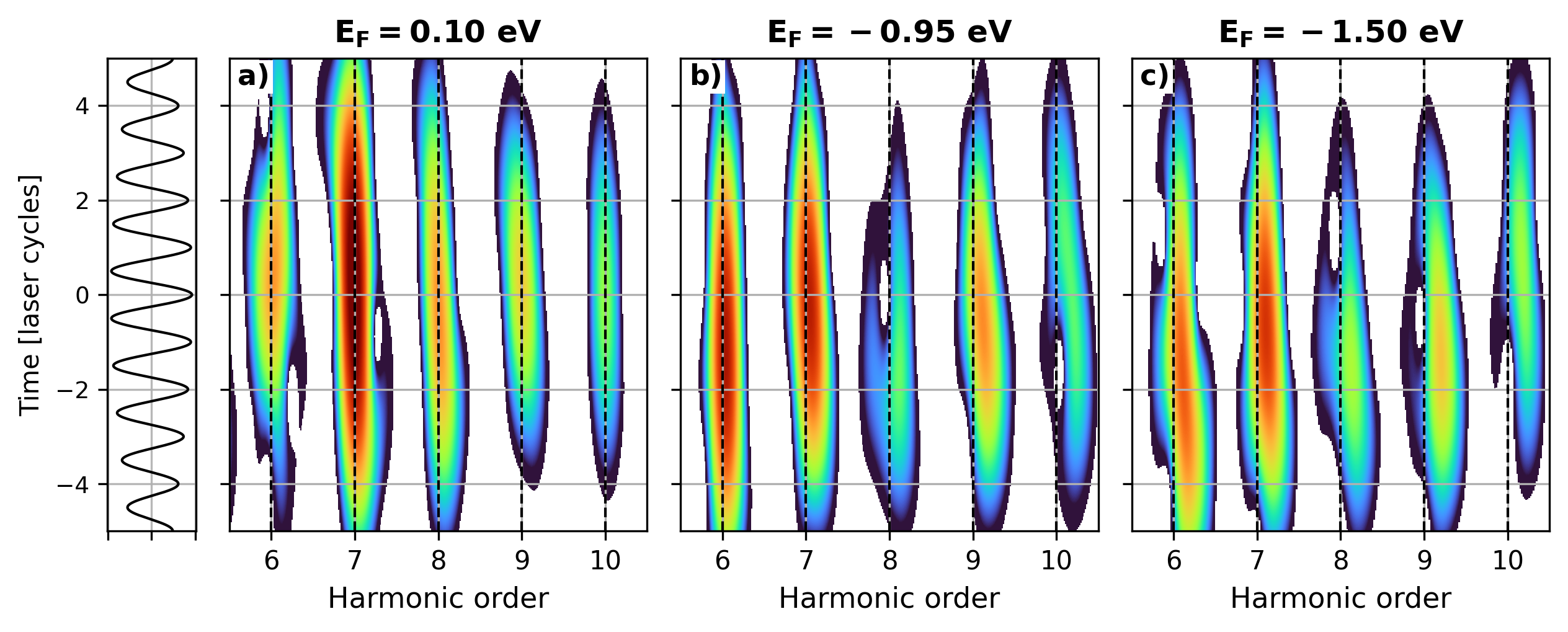}
\par\end{centering}
\caption{Time-frequency map of the currents obtained in Section~\ref{subsec:HHG} for NbSe$_2$ using a Gaussian window with variance $\alpha=(0.1\,\omega)^{-1}$. Three different values of the Fermi level are used: (a) $E_F=$~0.10, (b) -0.95, and (c) -1.5 eV. \label{fig:blue_shift}}
\end{figure}

In Fig.\ref{fig:blue_shift} we plot the time-frequency map of the currents generated by the NbSe$_2$ system for three different values of the Fermi level. It shows how harmonics are emitted at earlier times (rising-edge of the pulse) as $E_F$ becomes smaller, which translates into the harmonic blue-shift observed in Fig.~\ref{fig: grow_EF}.

\bibliographystyle{apsrev4-1}
\bibliography{references}% Produces the bibliography via BibTeX.

\end{document}